\documentclass[11pt]{article}
\usepackage{cite}
\usepackage{latexsym}
\usepackage{amssymb}
\usepackage{epsf}
\usepackage{amsmath}
\usepackage[hypertex]{hyperref}
\usepackage{graphicx}

\usepackage{slashed}

\newcommand{\tr}{{\rm Tr}}

\begin{document}
\pagestyle{empty}

\begin{flushright}
TU-909
\end{flushright}


\begin{center}

{\bf\LARGE Hadron physics as Seiberg dual of QCD\footnote{Talk at the
 GUT 2012 workshop, Kyoto, Japan, March 15-17, 2012.}}
\\

\vspace*{1.cm}
{\large 
Ryuichiro Kitano
} \\
\vspace*{0.5cm}

{\it Department of Physics, Tohoku University, Sendai 980-8578, Japan}\\
\vspace*{0.5cm}

\end{center}


\begin{abstract}
We try to identify the light hadron world as the magnetic picture of
QCD. We take both phenomenological and theoretical approaches to this
hypothesis, and find that the interpretation seems to show interesting
consistencies. In particular, one can identify the $\rho$ and $\omega$
mesons as the magnetic gauge bosons, and the Higgs mechanism for them
provides a dual picture of the color confinement\footnote{This talk is
based on Refs.~\cite{Kitano:2011zk, Kitano:2012zz}, where
Ref.~\cite{Kitano:2012zz} is a collaboration with Mitsutoshi Nakamura
and Naoto Yokoi.}.
\end{abstract}



\section{Introduction}

Seiberg duality in ${\cal N}=1$ supersymmetric gauge theories relates a
low energy physics of strongly coupled theory to a weakly coupled one
with a different gauge group~\cite{Seiberg:1994pq}. Since this is a
strong-weak duality, it has been believed to describe the
electric-magnetic duality in non-abelian gauge theories.
When the magnetic gauge theory is a free theory at low energy, one can
use the perturbation theory for the analysis such as the study of the
vacuum structure. Of particular interest is a vacuum where the magnetic
gauge group is Higgsed. The Higgsing is supposed to describe the color
confinement in the electric picture via the dual Meissner effect.

In this article, we try to approach to low energy QCD from a
supersymmetric QCD by using Seiberg duality.

\section{Evading the Vafa-Witten theorem}

In the ${\cal N}=1$ supersymmetric QCD with $N_c$ colors and $N_f$
flavors, the low energy description is an IR free gauge theory when
$N_c + 1 < N_f \leq 3N_c/2$. The gauge group in the magnetic theory is
$SU(N_f - N_c)$, and $N_f$ flavors of quarks and $N_f \times N_f$ mesons
are coupled.
By deforming these theories by adding small soft supersymmetry breaking
terms, one can obtain non-supersymmetric theories while maintaining the
viability of dualities.

Aharony {\it et~al.} first studied such deformations and found various
interesting vacua which may or may not be connected to the QCD
vacuum~\cite{Aharony:1995zh}.
Later, Arkani-Hamed and Rattazzi have established relations among the
soft terms in electric and magnetic
descriptions~\cite{ArkaniHamed:1998wc}. In particular, they found in the
free magnetic range that the origin of the moduli space is unstable, and
at least a vector-like symmetry, $SU(N_f)_V$ or $U(1)_B$, must be
spontaneously broken when the soft terms are small.
This results immediately tell us that there must be a phase transition
as we make the soft terms large, since there is a general theorem which
forbids a spontaneous breaking of vector-like symmetries in
QCD~\cite{Vafa:1983tf}.

However, there is a simple way to evade the theorem. The problem above
arises from the fact that the dual squark direction of the potential is
unstable. Therefore, if the dual squarks do not carry the baryon number,
the $U(1)_B$ symmetry is left unbroken and the theorem does not forbid
the smooth connection to a regime of large soft terms.

Our trick to make dual squarks neutral under $U(1)_B$ is to split
flavors into $N_f + N_c$~\cite{Kitano:2011zk}.
We start with $SU(N_c)$ gauge theory with $N_f + N_c$ flavors and add
masses to $N_c$ flavors. The massive flavors can be thought of as a
regulator which we are not interested in, but necessary to include it to
detour the phase transition.
The quantum numbers of the model are listed in
Table.~\ref{tab:ele-K}. The chiral superfields $Q$ and $\bar Q$ are the
massless quarks and $Q^\prime$ and $\bar Q^\prime$ are massive
regulators. Note that $U(1)_B$ is defined so that $Q$ and $\bar Q$ have
charges $1$ and $-1$, respectively, and the regulators are neutral.
The general theorem says that the $U(1)_B$ should not be broken, whereas
$U(1)_{B^\prime}$ can be broken spontaneously since one can simply gauge
it (or break it explicitly) without changing the physics of the QCD
sector.

\renewcommand{\arraystretch}{1.3}
\begin{table}[t]
\small
\begin{center}
 \begin{tabular}[t]{cccccccc}
& $SU(N_c)$ & $SU(N_f)_L$ & $SU(N_f)_R$ & $U(1)_B$ 
& $SU(N_c)_V$ &  $U(1)_{B^\prime}$ & $U(1)_R$ 
\\ \hline
 $Q$ & $N_c$ & $N_f$ & $1$ & 1 & 1 & 0 &
$(N_f - N_c)/N_f$ \\
 $\overline Q$ & $\overline {N_c}$& 1 & $\overline {N_f}$ & $-1$ 
& 1 & 0 & $(N_f - N_c)/N_f$ \\ \hline
 $Q^\prime$ & $N_c$ & 1 & $1$ & 0 & $\overline{N_c}$ & 1 &
1 \\
 $\overline Q^\prime$ & $\overline {N_c}$& 1 & 1 & 0 
& ${N_c}$ & $-1$ & 1 \\ 
 \end{tabular}
\caption{Quantum numbers in the electric picture.}
\label{tab:ele-K}
\end{center}
\end{table}
\renewcommand{\arraystretch}{1}

For $N_c + 1 < N_f + N_c \leq 3N_c/2$, {\it i.e.,} $1 < N_f \leq N_c/2$,
there is a dual magnetic picture which is IR free. For $N_c/2 < N_f <
N_c$, which is the case for two-flavor QCD, the both electric and
magnetic pictures are in the conformal window, and the magnetic picture
is more weakly coupled.
The particle content and the quantum numbers in the magnetic picture are
listed in Table.~\ref{tab:mag-K}.
As one can see, a part of the squarks, $q$ and $\bar q$, are neutral
under $U(1)_B$, and thus a vacuum with non-vanishing vacuum expectation
values for $q$ and $\bar q$ can be smoothly connected to the
non-supersymmetric limit.

We gauge the $U(1)_{B^\prime}$ symmetry to avoid the appearance of the
massless mode. As stated before, the physics of the QCD sector is
unchanged by this gauging since only the regulator fields couple to this
gauge boson.
The potential of the model was studied in this setup, and indeed a
minimum is found at $q = \bar q \propto {\bf 1}_{N_f \times N_f}$ and
$q^\prime = \bar q^\prime =0$ in a wide region of the parameter
space~\cite{Kitano:2011zk}.

\renewcommand{\arraystretch}{1.3}
\begin{table}[t]
\begin{center}
\hspace*{-.8cm}
\small
 \begin{tabular}[t]{ccccccccc}

& $SU(N_f)$ & $SU(N_f)_L$ & $SU(N_f)_R$ & $U(1)_B$ 
& $SU(N_c)_V$  & $U(1)_{B^\prime}$ & $U(1)_R$ 
\\ \hline
 $q$ & $N_f$ & $\overline{N_f}$ & 1 & 0 
& 1 & $N_c/N_f$ &  $N_c/N_f$ \\ 
 $\overline q$ & $\overline {N_f}$ & 1 & $N_f$ & 0 
& 1 & $-N_c/N_f$ &  $N_c/N_f$ \\ 
 $\Phi$ & 1 & $N_f$ & $\overline{N_f}$ & 0 
& 1 & 0 &  $2 (N_f - N_c)/N_f$ \\ \hline
 $q^\prime$ & $N_f$ & 1 & 1 & 1 
& ${{N_c}}$ & $-1 +  N_c/N_f$ &  0 \\ 
 $\overline q^\prime$ & $\overline {N_f}$ & 1 & 1 & $-1$ 
& $\overline{N_c}$ & $1 - N_c/N_f$ &  0 \\ 
 $Y$ & 1 & 1 & $1$ & 0 & 1 + Adj. & 0 &
2 \\
 $Z$  & 1 & 1 & $\overline {N_f}$ & $-1$ 
& $\overline{N_c}$ & 1 & $(2 N_f - N_c)/N_f$ \\ 
 $\overline Z$ & 1 & $N_f$ & 1 & 1 & ${N_c}$ & $-1$ &
$(2 N_f - N_c)/N_f$ \\

 \end{tabular}
\caption{Quantum numbers in the magnetic picture.}
\label{tab:mag-K}
\end{center}
\end{table}
\renewcommand{\arraystretch}{1}

\section{Hidden Local Symmetry}

The vacuum we find in the deformed supersymmetric QCD is very similar to
the one in QCD.
The condensation of the dual squarks breaks $SU(N_f)_L \times SU(N_f)_R$
down to $SU(N_f)_V$, providing massless pions.
Also, the magnetic gauge group $SU(N_f)$ is completely Higgsed and
locked to $SU(N_f)_V$. The massive magnetic gauge bosons, therefore,
transform as the adjoint representation under $SU(N_f)_V$. These gauge
bosons can naturally be identified as the vector mesons in QCD. 
Indeed, it has been known that the picture of identifying the vector
mesons as the gauge bosons of $SU(N_f)_V$ is phenomenologically very
successful~\cite{Bando:1984ej}.
The lowest level Lagrangian based on the Yang-Mills theory is called the
``hidden local symmetry,'' which has the same structure as the magnetic
model we obtained. (See Refs.~\cite{Seiberg:1995ac, Harada:1999zj,
Komargodski:2010mc, Abel:2012un} for discussion on the interpretation of
vector mesons as the dual gauge bosons.)

We argue that the similarity of the low energy physics supports a smooth
connection between the supersymmetric QCD with $N_f + N_c$ flavors and
non-supersymmetric QCD with $N_f$ flavors.

\section{Confining string as a meson vortex}

We have seen that the vector mesons may be identified as the magnetic
gauge bosons of QCD. If this is the case, the Higgs mechanism for the
vector meson masses should describe the color confinement in the
electric picture. A vortex of the vector mesons, which can be
constructed classically, corresponds to the confining string.
We here check this hypothesis quantitatively~\cite{Kitano:2012zz}.

We start with the Lagrangian as follows:
\begin{eqnarray}
 {\cal L}
&=& -{1 \over 4} F_{\mu \nu}^{(\omega)} F^{(\omega) \mu \nu}
 -{1 \over 4} F_{\mu \nu}^{(\rho)a} F^{(\rho) \mu \nu a}
\nonumber \\
&&
+ {f_\pi^2 \over 2}
\tr \left[
|D_\mu H_L|^2 + |D_\mu H_R|^2
\right]
\nonumber \\
&& - V(H_L, H_R).
\label{eq:lag}
\end{eqnarray}
This Lagrangian is meant to be a low energy effective theory of vector
mesons and pions. We have introduced $\rho$ and $\omega$ mesons as
$U(N_f)$ gauge group, and the Higgs fields $H_L$ and $H_R$ (correspond
to $q$ and $\bar q$ in the previous section) transform as $(N_f, \bar
N_f, 1)$ and $(1, N_f, \bar N_f)$ under $(U(N_f)_L, U(N_f), U(N_f)_R)$
groups, where $U(N_f)_L \times U(N_f)_R$ is the chiral symmetry, and
$U(N_f)$ in the middle is the gauge group.

The potential term $V$ is arranged so that it has a minimum at $H_L =
H_R = {\bf 1}$ where the chiral symmetry is spontaneously broken and
massless pions appear. Simultaneously, the $\rho$ and $\omega$ mesons
obtain masses. There are also massive scalar fields (the Higgs
bosons). We identify the $CP$-even Higgs fields as the scalar mesons
$f_0$ and $a_0$.

The parameter of the model can be fixed by the meson masses and
couplings. In particular, the gauge coupling constant can be extracted
from the decay constant of the $\rho$ and $\omega$ mesons:
\begin{eqnarray}
 g = {m_\rho^2 \over g_\rho} \simeq 5.
\end{eqnarray}
With this gauge coupling constant and meson masses, all the parameters
of the model relevant for the discussion are fixed.

\begin{figure}[t]
\begin{center}
 \includegraphics[width=10cm]{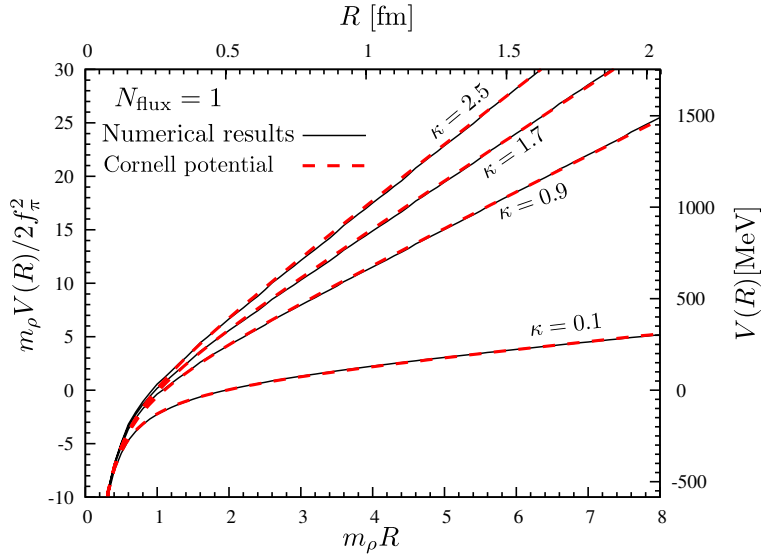}
\caption{Potential energy of the monopole-antimonopole system for 
$\kappa = 0.1$, 0.9, 1.7 and 2.5. The fittings with the 
Cornell potential are superimposed (dashed lines).}
\label{fig:energy}
\end{center}
\end{figure}

One can construct a classical vortex configuration of the string by
solving the equations of motion while imposing suitable boundary
conditions.
By using the configurations of scalar and gauge fields, the potential
energy of the string can be calculated.
We show in Fig.~\ref{fig:energy} the potential energies of string
configurations where we have used Dirac monopoles and anti-monopoles as
the boundary conditions. 
The magnetic charge is taken to be the minimal one allowed by the Dirac
quantization condition.
The separation of the monopole and the anti-monopole is taken to be
$R$. The $\kappa$ parameter is a ratio between the scalar and vector
mesons: $\kappa = {m_{a_0} / \sqrt{2} m_\rho}$, where the QCD data
suggests $\kappa \sim 0.9$.
The numerically obtained potential can be well fitted by the Cornell
potential:
\begin{eqnarray}
 V = - {A \over R} + \sigma R,
\label{eq:cornell}
\end{eqnarray}
with 
\begin{eqnarray}
 A \simeq 0.25,\ \ \ \sqrt \sigma \simeq 400~{\rm MeV}.
\label{eq:value}
\end{eqnarray}

The quarkonium spectrum and the lattice calculations support the shape
of the potential in Eq.~\eqref{eq:cornell} with values:
\begin{eqnarray}
 A = 0.25 - 0.4,\ \ \ \sqrt \sigma \simeq 430~{\rm MeV}.
\end{eqnarray}
These numbers are consistent with the ones in Eq.~\eqref{eq:value}.

\section{Conclusions}

Dualities provide non-perturbative connections among theories which are
defined differently. In this network of theories, there can be various
paths to approach to QCD.
We have seen that a non-supersymmetric deformation of the supersymmetric
$SU(N_c)$ gauge theory with $N_f + N_c$ flavors is a good candidate. If
this is the case, one can understand the confinement and chiral symmetry
breaking of QCD as condensations of the dual squarks, and the massive
vector mesons $\rho$ and $\omega$ are identified as the magnetic gauge
bosons.

If there is a smooth path from supersymmetric theory to QCD, it suggests
that there is a non-abelian electric-magnetic duality in QCD.
In this case, a lattice simulation should be able to see an interesting
phenomenon; the vector mesons become very light (namely $m_\rho \ll
m_{a_1}$) by an appropriate tuning of parameters, where supersymmetry
may not be important. The presence of such a parameter will be an
evidence for the existence of a dual gauge theory where the vector and
scalar mesons are the effective degrees of freedom.


\section*{Acknowledgements}
I would like to thank organizers of the GUT 2012 workshop. This work is
supported in part by the Grant-in-Aid for Scientific Research 23740165
of JSPS.





%
\end{document}